\documentclass[10pt,twocolumn]{article}
\usepackage[square, sort&compress, numbers]{natbib} % allows bibtex
\usepackage{multibib} % For SI refferences
\newcites{si}{Appendix References}
\usepackage{hyperref} % For \texorpdfstring command
\usepackage[inline]{enumitem} % allows inline lists (can easily be removed)
\usepackage{siunitx} % allows SI units
\usepackage{graphicx} % Allows images to be shown
\usepackage{xspace} % Space after command
\usepackage{amsfonts} % Allows for mathbb
\usepackage[fleqn]{amsmath} % Binomial coef. fleqn left aligns equations
\usepackage{authblk} % Allows for multiple authors
\usepackage{mathrsfs} % Allows for mathscr
\usepackage[margin=0.75in]{geometry} % Allows for margins/multicolumns
\usepackage{algorithm2e} % Allows for algorithms in SI
\usepackage{chemformula}
\usepackage[section]{placeins} 
\usepackage[shortcuts]{extdash} % hyphenate hyphenated words

\usepackage{xcolor} % allows us to create highlights

\usepackage{orcidlink}
\providecommand{\orcidlink}[1]{}
\usepackage{appendix}
\DeclareMathOperator\erf{erf}
\newcommand{\R}{\mathbb{R}}

\newcommand{\nm}[1]{\SI{#1}{\nano\meter}}
\newcommand{\fs}[1]{\SI{#1}{\femto\second}}
\newcommand{\ps}[1]{\SI{#1}{\pico\second}}
\newcommand{\K}[1]{\SI{#1}{\kelvin}}

\newcommand{\naive}{na\"{i}ve\xspace}
\newcommand{\vs}{\emph{vs.}\xspace} % make vs consistent

\newcommand{\siel}[1]{%
  \ifnum#1=0%
  \else%
    \ifnum#1=1%
      Si% no subscript
    \else%
      Si\textsubscript{#1}% add subscript
    \fi%
  \fi%
}
\newcommand{\hel}[1]{%
  \ifnum#1=0%
  \else%
    \ifnum#1=1%
      H% no subscript
    \else%
      H\textsubscript{#1}% add subscript
    \fi%
  \fi%
}
\newcommand{\sih}[4]{%
  \siel{#1}\hel{#2}{%
    \ifnum%
      \numexpr#3+#4>0 -%
    \fi%
  }\siel{#3}\hel{#4}\xspace%
}
\newcommand{\ie}{\emph{i.e.},\xspace}
\newcommand{\eg}{\emph{e.g.},\xspace}
\newcommand{\eq}[1]{(\ref{eq:#1})}

\newcommand{\sisec}[1]{Appendix Section~\ref{sisec:#1}}

\newcommand{\sinumbering}{
    \preto{\section}{%
      \setcounter{figure}{0} %
      \setcounter{table}{0} %
      \setcounter{equation}{0} %
      \setcounter{algocf}{0} %
    } %
    \renewcommand{\thefigure}{\thesection\arabic{figure}} %
    \renewcommand{\thetable}{\thesection\arabic{table}} %
    \renewcommand{\theequation}{\thesection\arabic{equation}} %
    \renewcommand{\thealgocf}{\thesection\arabic{algocf}} %
}

\newcommand{\Rs}{\mathbb{R}} % Reals
\newcommand{\set}[1]{\left\{#1\right\}} % set comprehension
\newcommand{\mc}[1]{\mathcal{#1}} % Calligraphic math
\newcommand{\subB}{\mathrm{B}}

\newcommand{\subst}{\mathrm{st}}

\def\mailto#1{{\tt #1}}
\def\ead#1{E-mail: \mailto{#1}}

\newcommand*{\pst}{$P_\subst$\xspace}

\newcommand{\ns}{m}
\newcommand{\nt}{n}
\newcommand{\hp}{\hat{p}}
\newcommand{\vf}{\vec{f}}
\newcommand{\bx}{\bar{\vx}}
\newcommand{\vx}{\vec{x}}

\renewcommand{\vec}[1]{\boldsymbol{#1}}

\renewcommand{\aa}{\textbf{a)}\xspace}
\newcommand{\bb}{\textbf{b)}\xspace}
\newcommand{\cc}{\textbf{c)}\xspace}
\newcommand{\dd}{\textbf{d)}\xspace}
\newcommand{\ee}{\textbf{e)}\xspace}
\newcommand{\ff}{\textbf{f)}\xspace}

\begin{document}

\title{Machine learning models for Si nanoparticle growth in nonthermal plasma}

\author[1]{Matt Raymond\orcidlink{0000-0001-6824-8692}}
\author[2]{Paolo Elvati\orcidlink{0000-0002-6882-6023}}
\author[3]{Jacob C. Saldinger\orcidlink{0000-0001-5005-614X}\footnote{Now at Low Carbon Pathway Innovation at BP}}
\author[1]{Jonathan Lin\orcidlink{0009-0004-6381-4068}}
\author[2]{Xuetao Shi\orcidlink{0000-0001-6274-5495}\footnote{Now at the Dana-Farber Cancer Institute at Harvard}}
\author[1,2,3]{Angela Violi\orcidlink{0000-0001-9517-668X}\thanks{\ead{avioli@umich.edu}}}

\affil[1]{Mechanical Engineering, University of Michigan, Ann Arbor, 48109-2125, Michigan, USA}
\affil[2]{Electrical Engineering and Computer Science, University of Michigan, Ann Arbor, 48109-2125, Michigan, USA}
\affil[3]{Chemical Engineering, University of Michigan, Ann Arbor, 48109-2125, Michigan, USA}

% % Double-blind peer review
% \author{
%     Anonymous Authors
% }

\maketitle

\begin{abstract}
Nanoparticles (NPs) formed in nonthermal plasmas (NTPs) can have unique properties and applications. 
However, modeling their growth in these environments presents significant challenges due to the non-equilibrium nature of NTPs, making them computationally expensive to describe. 
In this work, we address the challenges associated with  accelerating the estimation of parameters needed for these models. 
Specifically, we explore how different machine learning models can be tailored to improve prediction outcomes. 
We apply these methods to reactive classical molecular dynamics data, which capture the processes associated with colliding silane fragments in NTPs. 
These reactions exemplify processes where qualitative trends are clear, but their quantification is challenging, hard to generalize, and requires time-consuming simulations.
Our results demonstrate that good prediction performance can be achieved when appropriate loss functions are implemented and correct invariances are imposed.
While the diversity of molecules used in the training set is critical for accurate prediction, our findings indicate that only a fraction (15-25\%) of the energy and temperature sampling is required to achieve high levels of accuracy. This suggests a substantial reduction in computational effort is possible for similar systems.
\end{abstract}

\vspace{2pc}
% 3-7 keywords allowed

% \vspace{2pc}
% \noindent{Submitted to: \it{Plasma Sources Sci. Technol.}}

\section{Introduction}
Nonthermal plasmas (NTPs) are unique environments where low-temperature neutral species and ions coexist with high-temperature electrons.
For this reason, these systems have received considerable attention, especially for synthesizing particles and nanoparticles with significant tunability.
This flexibility in the final particle properties results from an environment with enough localized energy to cross relatively high free energy barriers while avoiding excessive thermal energy and discouraging agglomeration~\citep{mangolini2009}.
As a result, the synthesis of nanoparticles and thin films under these conditions holds potential applications in biomedicine~\citep{gao2004, fujioka2008}, energy~\citep{moore2011, saadane2003, dogan2016}, microelectronics~\citep{weis2011}, and catalysis~\citep{astruc2005}.

However, modeling these environments remains a significant challenge due to the combined non-equilibrium and multiscale nature.~\citep{boufendi1994, lanham2022, kruger2019}
Even when narrowing the description only to a specific scale or class of processes, such as particle growth (\eg nucleation, coagulation, surface deposition)~\citep{kortshagen1999, agarwal2014, lepicard2016}, the accuracy of the methods depends on their ability to model a variety of size- and charge-dependent growth mechanisms.
These processes, in turn, depend on the propensity of species, generally radicals, to form stable bonds upon collision with other particles or surfaces. 
Still, these processes have been frequently estimated using fixed values, independent of the colliding species and energies~\cite{lepicard2016}, primarily due to the complexity of obtaining a more detailed functional form.
Recently~\cite{shi2021}, we have shown how atomistic simulations can capture the complex reactivity of small neutrals and provide parameters that can be used in reactor models~\cite{lanham2021, husmann2023}.
While these previous and current works focus on silane particles, the underlying methodology is general and adaptable to various conditions where species' internal and translational energy distributions differ. 
This flexibility sets our method apart from others, such as the one recently published by Bal and Neyts~\cite{bal2021}, which (among other differences) does not make assumptions about the translational energy distribution or the specific reaction under investigation. 
Due to the large size of some involved species and the resulting (lack of) separation of vibrational modes, we observed a variety of competing reaction mechanisms involving a complex interplay of physisorption, chemisorption, and desorption.

While very informative, deriving these reacting probabilities via molecular dynamics (MD) simulations remains time-consuming and computationally burdensome.
Even when using classical reactive MD, a timestep of the order \SI{10}{\atto\second} is required to guarantee correct numerical integration of the equations of motion during the reactions.
Moreover, due to numerous variables (\eg impact parameter, speed distribution, surface composition) and relevant species present in such reactive systems, the number of simulations required to capture the collisions experiences rapid combinatorial growth.
More effective means of deriving the collision parameters must be considered to scale this approach.
In this work, we focus on machine learning (ML) methods, which offer the potential to formulate a dependency between the system conditions and the final collision outcome. Data-driven methods do not remove the need for MD simulations but allow for a drastic reduction in computational effort.

Recent works have used ML methods to overcome similar combinatorial problems associated with the growth of nanoparticles in reactive gas-phase environments, accurately predicting the aggregation propensities of soot precursors~\cite{saldinger2023}.
However, no existing work addresses the same scientific questions in the context of nonthermal plasmas.
Most studies focus on predicting plasma properties~\citep{gideon2019, gaag2021, liang2023} and plasma-surface interactions, from surface deposition~\citep{han1994, guessasma2004, pakseresht2015, kruger2019, gergs2023}, plasma etching~\citep{hong2003, kim1994, kwon2022}, and surface modification~\citep{abdjelil2013}.
In contrast, this work examines a scale between detailed individual reactions and simplified larger systems, where detailed chemistry must be approximated. Our focus is on particles approximately \nm{1} in diameter colliding with small reactive fragments (\ch{SiH_y} and \ch{Si2H_y}).
Building upon previous data~\cite{shi2021}, we demonstrate how easily computable properties can be used to train ML models to generate predictions for new species or mitigate the MD computational cost.

\section{Methodology}
\subsection{Molecular Dynamics Simulations}
We performed classical reactive molecular dynamics simulations to study the collisions between disilanes, \ch{Si2Hx}, and other silane clusters and molecules using the same procedure described previously~\cite{shi2021}.
First, we independently equilibrated both colliding species' rotational and vibrational modes, and then we performed microcanonical simulations at a fixed impact velocity.
Between 40 and 100 different collision vectors were imposed to parallel the line passing through the center of mass of the two species (\ie impact parameter = 0 or, equivalently, impact angle = $\pi$).
Simulations were performed using LAMMPS~\cite{plimpton1995} using the ReaxFF force field~\cite{rappe1991} in combination with a dynamic charge equilibration model~\cite{nakano1997} and integrated the equations of motion every \fs{0.01}.
To analyze the collision outcome, we monitored the minimum distance between all the atoms or only the Si atoms of each cluster.

The conformations of the colliding species were generated from the canonical simulations at \K{300}, \K{400}, \K{500}, \K{600}, and \K{900}.
Properties are computed by reweighting the collisions using a Maxwell-Boltzmann distribution, which allows labeling each system with a single temperature.
For clarity, we grouped the colliding species in two sets, labeled ``clusters'' and ``impactors'', but there is no physically meaningful distinction associated with each set.
Molecules in the cluster set are silanes that cover different sizes and H-coverage (\ie \ch{Si2H6,\ Si4}, and \ch{Si29H_x} with $x=18,27,31,36$), while as impactors, we considered different disilanes (\ie \ch{Si2H_x} with $x \in [1,6]$) in all possible hydrogen distribution (\eg for \ch{Si2H4}, we simulated both \ch{H2Si^.-^.Si}\ch{H2} and \ch{HSi^{..}-SiH3}).
The results of these simulations were combined with previously computed data~\cite{shi2021} to create a dataset of 390 collision pairs based on approximately 650\,000 simulations.

\subsection{Machine Learning}
We compared the predictive performances of seven standard ML models for predicting sticking probabilities: an unpenalized linear model, ElasticNet, Kernel Ridge Regression (KRR), Support Vector Regression (SVR), $k$-nearest neighbors (KNN)~\citep{pedregosa2011}, DeepSets~\citep{zaheer2017}, and Light Gradient\-/Boosting Machine (LGBM)~\citep{ke2017}.

\subsubsection{Input features}
For model input, we generated feature vectors using parameters that describe properties likely to affect the sticking probability for silane molecules~\citep{shi2021} (\ie H coverage, temperature, and molecule's size). 
Specifically, for each cluster and impactor, we used the number of Si atoms, H atoms, and a vector of the number of unpaired electrons per Si atom (to differentiate between isomers).
For the disilanes, this two-dimensional vector indicates the number of unpaired electrons on each Si atom; in contrast, only the total number was used for the larger clusters, and the second element was always set to 0.
For particle $a$, we denote this feature vector as $\vf_a \in \Rs^4$.
Each particle interaction has an associated translational temperature, denoted $t \in (0,\infty)$.
Thus, we denote the feature vector for a pair of particles as $\vx_{a,b} \doteq [\vf_a^\top \; \vf_b^\top \; t]^\top \in \Rs^9$.
These nine features were selected because they are computed efficiently and were expected to capture much of the relevant chemistry.
We normalized each training, validation, and testing dataset so that the concatenation of the training and validation sets has a mean of 0 and a standard deviation of 1 for each feature.

\subsubsection{Loss functions}
Each plasma simulation can be interpreted as a binomial distribution since each outcome is a Bernoulli trial for some probability $p$.
We use the negative log-likelihood of the binomial distribution (B-NLL) as a loss function for a given simulation,
\begin{equation}
    \ell_{b}(\hp, \ns, \nt) \doteq - [\ns \log \hp + (\nt - \ns) \log (1-\hp)] \;,
\end{equation}
where $\hp$ is the predicted probability, $\ns$ is the number of events for the desired outcome (\eg sticking), $\nt$ is the total number of events in a simulation.

We implement DeepSets with a sigmoidal activation $\sigma(u) \doteq (1+e^{-u})^{-1}$ on the output layer where $\sigma\colon \Rs \to [0,1]$ and directly optimize the B-NLL loss.
Unfortunately, many ML libraries do not natively support binomial loss functions.
In such cases, we can rewrite the binomial loss $\ell_{b}$ in terms of the logistic loss $\ell_l$,
\begin{equation}
\begin{split}
    \ell_{b}(\hp, \ns, \nt) = - \left[ \ns \ell_l(\hp, 1) + (\nt - \ns)\ell_l(\hp, 0) \right]\\
    \text{for} \quad \ell_l (\hp, b) \doteq \begin{cases}
        \log \hp & b=1 \\
        \log (1-\hp) & b=0
    \end{cases} \;,
\end{split}
\end{equation}
where $b$ indicates a class label.
This can be interpreted as logistic regression that includes both classes for each set of trials but weights each ``pseudosample'' according to the number of positive and negative events.
We use this approach for Logistic ElasticNet and LGBM.

\begin{figure}
    \centering
    \includegraphics{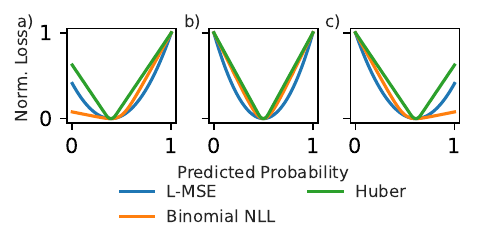}
    \caption{
        B-NLL, L-MSE, and logit-transformed Huber (L-H) losses rescaled and aligned for true {\pst}s of \aa 0.1, \bb 0.5, \cc 0.9.
    }
    \label{fig:binom_vs_mse}
\end{figure}

Another perspective is to interpret the event probability as a scalar $p\in [0,1]$ and perform regression in logit-space.
We cannot perform unconstrained regression directly on probabilities, as the model may predict unphysical values outside $[0,1]$.
Instead, we apply the logistic unit (logit) $\sigma^{-1}(p) \doteq \log (1/(1-p))$ where $\sigma^{-1}\colon [0,1] \to \Rs$ to restrict the (untransformed) output range.
Note that $\{0,1\}$ values cannot be predicted with a finite model output when using logits, so we clip the true probabilities at $[\epsilon,1-\epsilon]$ for some small $\epsilon$.
We refer to this loss as the ``Logit MSE'' (L-MSE).
To further emulate the binomial NLL, we weigh the loss for each simulation according to the number of trials and refer to it as the ``Logit-Weighted MSE'' (LW-MSE).
The LW-MSE penalizes outliers more significantly than Binomial NLL, as seen in Figure \ref{fig:binom_vs_mse}.
We use the LW-MSE for the unpenalized linear model, ElasticNet, KRR, and SVR, and also evaluate it on DeepSets and LGBM.

KNN does not utilize a loss function and is automatically restricted to the range $[0,1]$ as predictions are a weighted average of training data points, weighted by distance.
We use a ``\naive'' predictor as a baseline, which predicts the mean of the training probabilities.

% Permutation invariant distance function
\subsubsection{Permutation invariance}
Because our two-particle systems are permutation invariant, we train permutation invariant models using either data manipulation or model construction.

Some ML model implementations cannot be customized to be permutation invariant by construction, so we must adjust the dataset rather than the model itself.
For linear models such as OLS and ElasticNet, permutation invariance can easily be achieved by defining $\bx_{a,b} \doteq [(\vf_a^\top + \vf_b^\top)/2\; t]^\top \in \Rs^{5}$, which is equivalent to having equal model weights for the same indices of $\vf_a,\vf_b$.
Crucially, this averaging approach is only valid for linear models and would reduce the expressiveness of nonlinear models such as LGBM.
Instead, for LGBM, we augment the dataset to contain $\vx_{a,b}$ and $\vx_{b,a}$.
Although helpful, this approach does not guarantee invariance, so we refer to it as pseudo-permutation invariant.
In both cases, normalization is applied after transforming the training and validation feature vectors.

Other models can be directly modified to learn permutation invariant functions.
KNN, KRR, and SVR all use distance metrics to express new points as combinations of training data points;
KNN directly uses a distance metric to select and weight neighbors, and KRR and SVR use the RBF kernel
\begin{equation}
    k(\vx_{a,b}, \vx_{c,d}) \doteq \text{exp}\left( -\gamma d(\vx_{a,b}, \vx_{c,d}) \right)
\end{equation}
for some distance metric $d: \R^9\times \R^9 \to [0,\infty)$ via the representor theorem.
For all three methods, we use the permutation invariant distance metric
\begin{equation}
    d(\vx_{a,b}, \vx_{c,d}) \doteq \min\left( \left\| \begin{bmatrix} \vf_a - \vf_c \\ \vf_b - \vf_d \\ t \end{bmatrix}\right\|_2^2 \begin{matrix} \\\\,\end{matrix}  \left\| \begin{bmatrix} \vf_a - \vf_d \\ \vf_b - \vf_c \\ t \end{bmatrix}\right\|_2^2 \right) \;.  
\end{equation}
This value can be interpreted as the minimum distance across all particle permutations within $\vx_{a,b}$ and $\vx_{c,d}$.

Finally, we use the DeepSets~\citep{zaheer2017} neural network (NN) architecture, which has the form 
\begin{equation}
    g\left(\begin{bmatrix}\vf_a \\ \vf_b \\ t\end{bmatrix}\right) \doteq \phi_\eta\left(
\begin{bmatrix} \rho\left(\psi_\theta(\vf_a), \psi_\theta(\vf_b)\right) \\ t\end{bmatrix}\right) \;,
\end{equation}
where $\phi_\eta\colon \R^4 \to \R^h, \psi_\theta:\R^{h+1} \to (0,1)$ are neural networks with a hidden width of $h$ and parameters $\eta, \theta$ and $\rho: \R^{2h} \to \R^h$ is a feature-wise mean or max.
This architecture is permutation invariant by construction and has been used for similar particle interaction problems~\citep{saldinger2023}.
Together, these approaches make our predictions \mbox{(pseudo-)}permutation invariant, which improves generalization capabilities.

\subsubsection{Cross-validation}
To estimate the models' performance under different scenarios, we considered multiple cross-validation (CV) techniques: 5-fold, leave\-/one\-/temperature\-/out, leave\-/one\-/impactor\-/out, and leave\-/one\-/cluster\-/out.
Furthermore, we selected the model parameters using a grid search to reduce human bias.
Notably, estimating model performance and performing model selection using the same cross-validation splits is known to overestimate performance and lead to biased models~\citep{varma2006, cawley2010}.
To combat such bias, we estimated the model performance using ``nested cross-validation.''
This approach estimates model performance using an outer CV loop, splits each training set into an inner CV loop, and uses the inner loop to select optimal hyperparameters for each outer fold.
Specifically, for each outer fold, we select the parameters with the lowest average loss for the inner test datasets and report the loss of the outer test set using these parameters.
This process, known as ``nested CV'' provides an almost unbiased estimate of the true error~\citep{varma2006}.

The inner CV loop was conducted similarly to the outer loop, with the outer training set being split for the inner CV loop.
For the 5-fold CV, we perform the inner CV using another random 5-fold CV.
We also perform leave-one-out CV on the inner loop for leave\-/one\-/temperature\-/ and leave\-/one\-/cluster\-/out CV.
However, because there were so many impactors, the inner fold was created by partitioning the training impactors into five folds, each containing multiple impactors.
In short, the inner loop for parameter selection uses the same split criteria as the outer loop; our preliminary tests show that this approach improves generalization to unseen clusters and impactors.

We use a modified CV approach to provide train, validation, and test sets.
The data is typically split into training and testing sets of relative size $k-1$ and $1$.
However, since the NN and LGBM utilize validation sets, we split our inner training dataset for these models into $k-2$ folds for training and one fold for validation.
If a model (\eg linear regression) didn't use a validation set, we combined the training and validation sets.

An algorithmic representation is shown in the Supplemental Materials (SI Algorithms~\ref{sialg:cv_nested} and \ref{sialg:cv_group}), while tables of the grid-searched parameters and values are included in \sisec{gridsearch_params}.
We estimate the performance standard deviation by performing nested CV with five random seeds (random seeds were shared when splitting the dataset and initializing the model states).
All other parameters were set to the library defaults.

\section{Results and Discussion}
\subsection{Molecular Simulation}
\begin{figure}[!htb]
  \centering
  \includegraphics[width=0.49\textwidth]{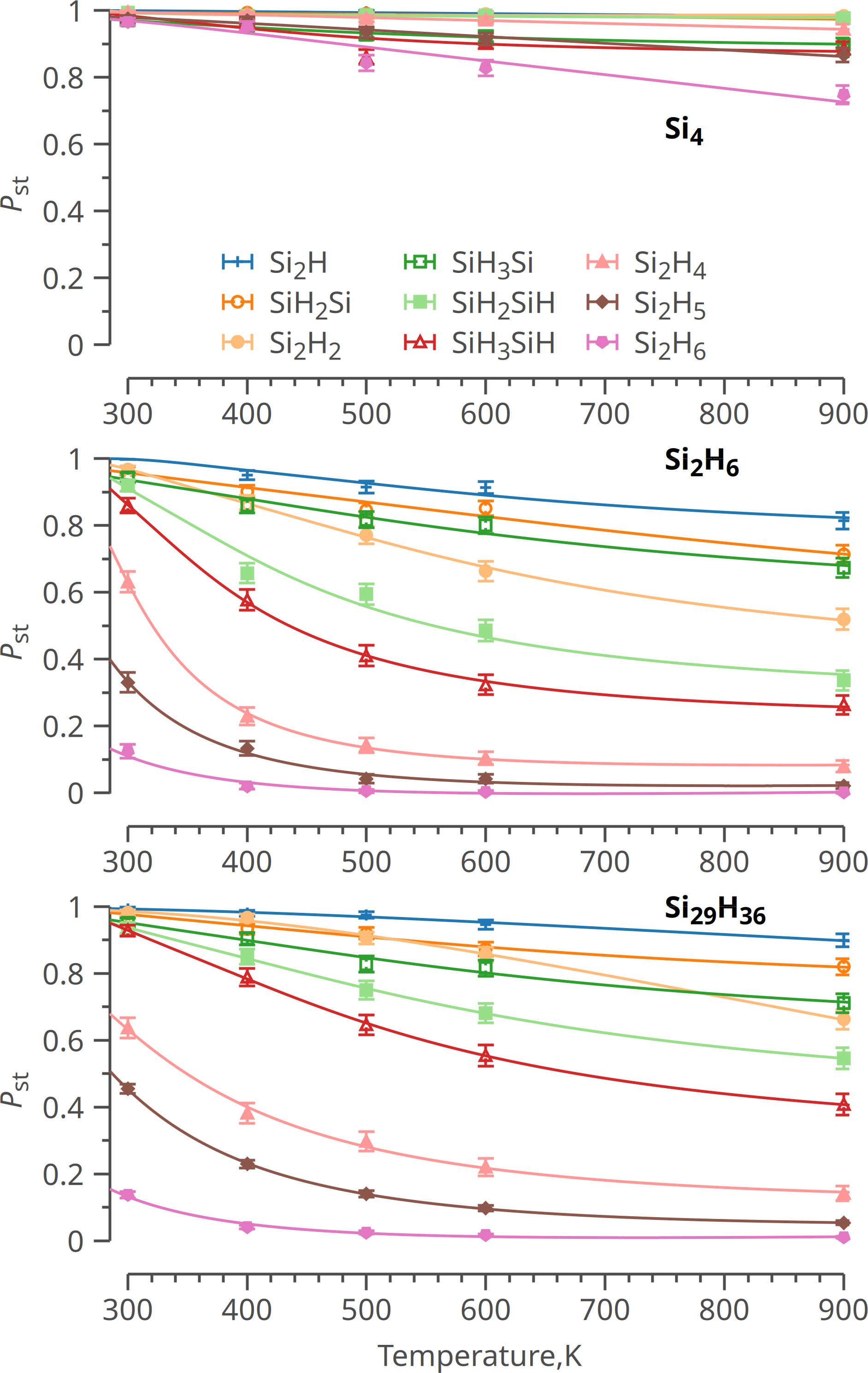}
  \caption{%\hl{\textbf{REDO}}
    Temperature dependence of the sticking probability for collisions between different \ch{Si2H_y} and \textbf{a)} \ch{Si4}, \textbf{b)} \ch{Si2H6}, and \textbf{c)} \ch{Si29H36}.
    Lines show the fitted trend, described in \eq{fit}. 
    Error bars represent two standard deviations.
    \ch{Si2H_y} indicate molecules with balanced hydrogen distribution, while unbalanced fragments are expanded for clarity. 
  }
  \label{fig:stickingsim}
\end{figure}

Previous work has analyzed the collisions of \ch{SiHx} as a critical step for the growth of particles in silane plasma~\citep{shi2021}. 
However, larger silanes also play a role in chemical growth despite decreasing concentration.
Figure~\ref{fig:stickingsim} shows \pst, the probability of a chemisorption or sticking event, for the collision of \ch{Si2H_x} with three of the simulated clusters.

Similarly to previous work~\citep{shi2021}, the trends of \pst are well approximated by
\begin{equation}
  m(T,E,b,c) = (1-c)f(T-b,E)+c \quad ,
  \label{eq:fit}
\end{equation}
where $T$ is the translational temperature, $E$ is the kinetic energy, $b$ and $c$ are fitted constants, and $f$ is the cumulative distribution function of the Maxwell-Boltzmann distribution:
\begin{equation}
  f(T,E)=\erf \left(\sqrt{\frac{E}{k_{\subB}T}}\right)-\frac{2}{\sqrt{\pi}}\sqrt{\frac{E}{k_{\subB}T}}\exp\left(-\frac{E}{k_{\subB}T}\right)
  \label{eq:cdf}
\end{equation}
in which $k_{\subB}$ is the Boltzmann constant and $\erf$ is the error function.

Compared to the results of \ch{SiH_y} collisions, the \pst for \ch{Si2H_y}, while generally slightly higher, displays very similar trends.
As expected, the sticking probability decreases at higher temperatures, with a stronger dependency observed for species with a lower number of radical electrons due to their high reactivity.
As before, the number of unpaired radicals plays a crucial role in the reactivity. 
However, the picture is complicated by the effect of balanced \vs imbalanced hydrogens on the silane impactors.
While a hydrogen imbalance results in greater reactivity, this effect is secondary to the overall hydrogen coverage.
Outliers like \ch{Si2H2} do not follow the expected trend, displaying a lower-than-expected propensity to form bonds at higher temperatures.

Cluster size has a relatively small effect, mostly apparent when physisorption is relevant, whether as a step for the chemisorption or as a collision outcome.
By analyzing the ratio between collisions that lead to chemisorption and the chemisorption and physisorption events (see  Figure~\ref{sifig:stickingtype} in the Supplementary Material), we observe that physisorption plays an integral role in the kinetics for almost fully saturated species like \ch{Si2H5 and \; Si2H4} isomers.
These reactants, which are less reactive than the more unsaturated counterparts or require as much energy as the fully saturated silanes, are the most likely to control the kinetics of particle growth.
It is worth noting that the lifetime of a physisorbed pair can vary from a few to tens of \ps{}, even when the outcome is a chemisorption.
As a result, in an experimental setting, other processes can occur in this timescale that the current model does not capture.

\begin{figure}[!htb]
\centering
  \includegraphics[width=0.49\textwidth]{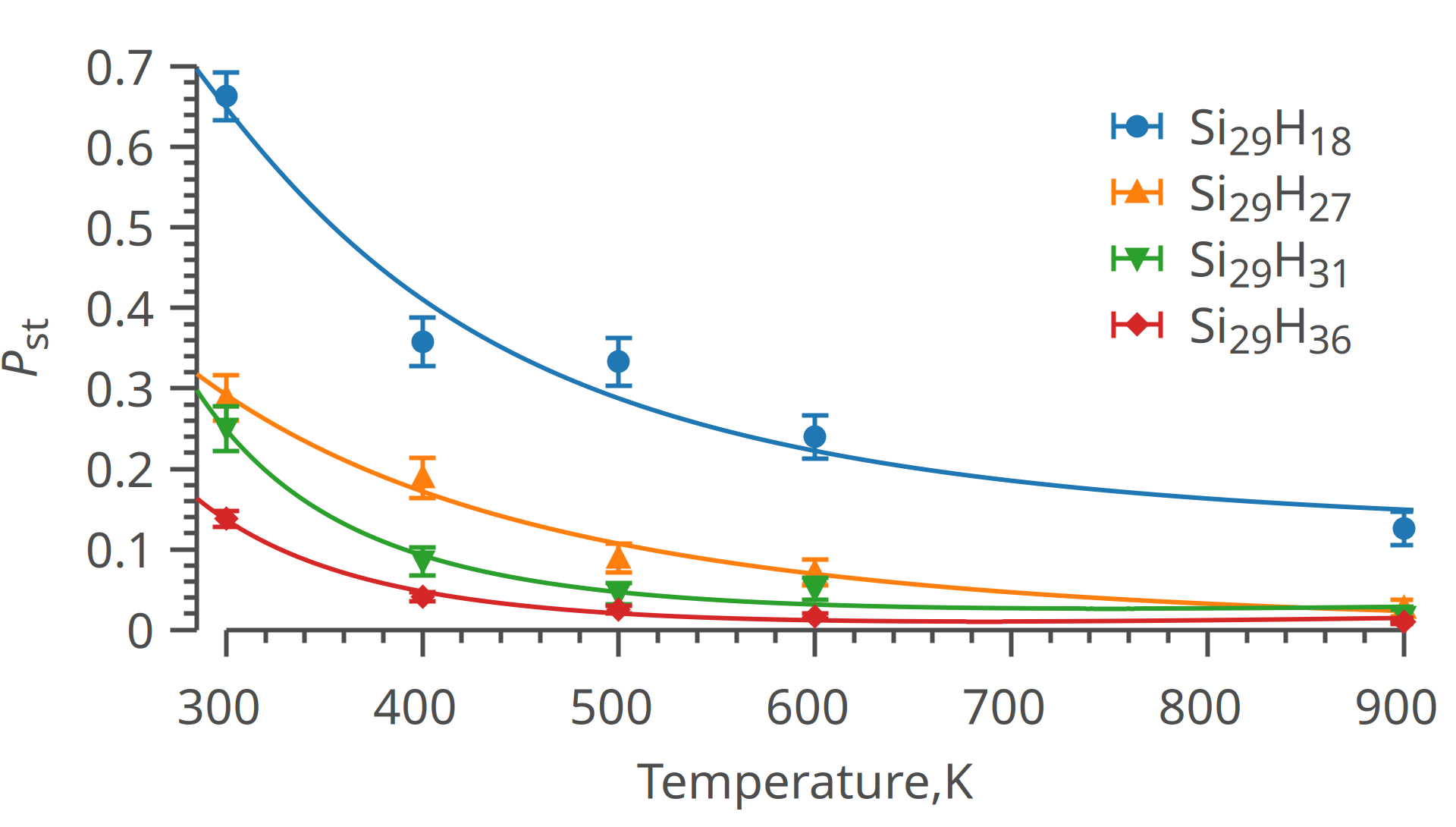}
  \caption{
    Sticking probability \vs temperature for the collisions between \ch{Si2H6} and \ch{Si29} cluster with different hydrogen coverages of \ch{Si29H_x}. 
    The line represents the fit discussed in \eq{fit}, and error bars (generally smaller than the symbols) represent two standard deviations.
  }
  \label{fig:stickingcoverage}
\end{figure}

Finally, our simulations also study \ch{Si29H_x} nanoparticles and hydrogen coverage to provide a quantitative relationship of the effect of hydrogen coverage of larger NPs on the sticking coefficients.
In Figure~\ref{fig:stickingcoverage}, we compare sticking coefficients with coverages of 18, 27, 31, and 36 hydrogens.
The same trends observed with other silane fragments are evident here: increasing hydrogen coverage while holding the temperature, impactor, and number of cluster silicon atoms monotonically increases the sticking probability.

\subsection{Machine learning models}
\label{sec:methods_ml}

\begin{figure}[!htb]
  \centering
  \includegraphics{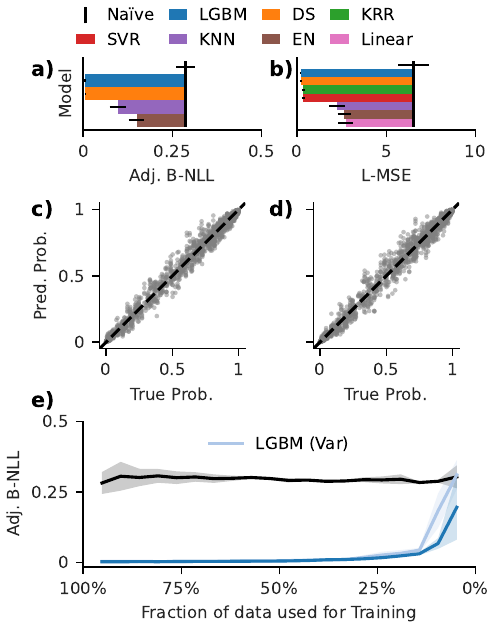}
  \caption{
    Performance of 5-fold cross-validation.
    \aa and \bb show the average performance of each model trained and evaluated using the adjusted B-NLL and L-MSE, respectively.
    Black bars indicate the standard deviation across all five folds and five random seeds.
    \cc and \dd show the LGBM predictions for all folds and seeds using the same loss functions as \aa and \bb, respectively.
    \ee shows the adjusted B-NLL performance of the permutation invariant (dark blue) and variant (light blue) LGBM models and the \naive model as the fraction of training data decreases.
    The shaded region indicates the standard deviation across random seeds.
  }
  \label{fig:5fold}
\end{figure}

\begin{figure}[!htb]
  \centering
  \includegraphics{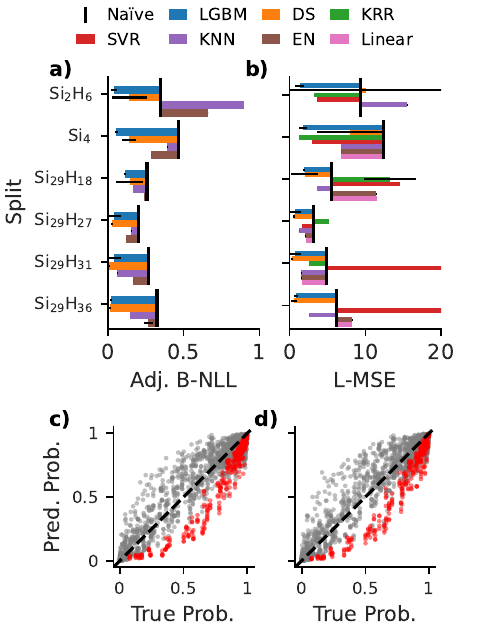}
  \caption{
    Performance of leave\-/one\-/cluster\-/out CV.
    \aa and \bb show the average performance of each model trained and evaluated using the adjusted B-NLL and L-MSE, respectively.
    Black bars indicate the standard deviation across random seeds.
    \cc and \dd show the predictions of LGBM for all folds and random seeds using the same loss functions as \aa and \bb, respectively.
    Predictions for \ch{Si29H18} are highlighted in red.
    The losses for SVR \ch{Si29H31} and \ch{Si29H36} in \bb are 23.5 and 41.5 but are truncated for visualization purposes.
  }
  \label{fig:leaveclusterout}
\end{figure}

\begin{figure}[!htb]
  \centering
  \includegraphics{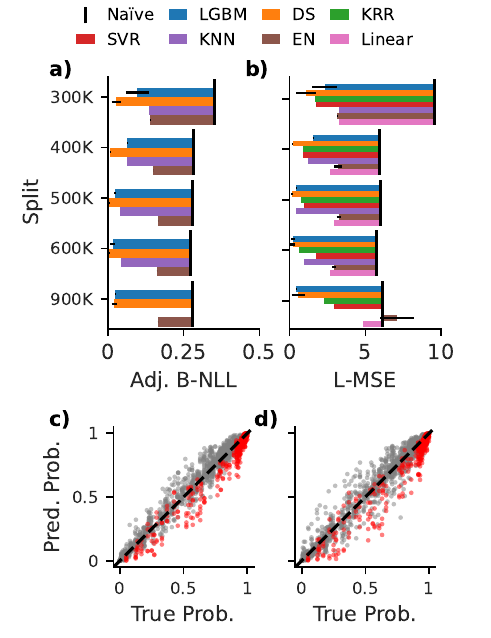}
  \caption{
    Performance of leave\-/one\-/temperature\-/out CV.
    \aa and \bb show the average performance of each model trained and evaluated using the adjusted B-NLL and L-MSE, respectively.
    \cc and \dd show the predictions of DeepSets for all folds and random seeds using the same loss functions as \aa and \bb, respectively.
    The predictions for \K{300} are highlighted in red.
  }
  \label{fig:leavetempout}
\end{figure}

As described in the Methodology, we performed cross-validations using several splits to determine which environmental settings our models could and could not generalize.
We test the overall capabilities of the model using a 5-fold CV (Figure~\ref{fig:5fold}) and out-of-distribution generalization by performing leave\-/one\-/temperature\-/, leave\-/one\-/impactor\-/, and leave\-/one\-/impactor\-/out CV (Figures \ref{fig:leavetempout} and \ref{fig:leaveclusterout}, and SI Figure~\ref{sifig:leaveimpactorout}).

The binomial NLL depends on the number of trials in each simulation and is nonzero for perfect predictions.
Furthermore, this nonzero floor is not constant and depends on the true probability.
Thus, comparing the binomial NLL between folds may be misleading, as different numbers of trial runs or distributions of actual probabilities may dominate variations.
To improve visualization, for plotting, we use the ``adjusted B-NLL'':
\begin{align}
    \ell_\text{adj}(\hp, m, n) \doteq \frac{1}{n}\left(\ell_b\left(\hp, m, n\right) - \ell_b\left(\frac{m}{n}, m, n\right)\right)
\end{align}
This loss weights all simulations equally, regardless of the number of trials run, and subtracts the NLL of a perfect prediction from the NLL of the actual prediction.
For similar reasons, we plot the unweighted L-MSE instead of the LW-MSE.
Unlike root mean squared error, these metrics do not correspond to intuitive notions of distance but still facilitate a quantitative performance comparison between methods.
Thus, we also plot the true \vs predicted probability for the sticking event to provide an intuition of how individual models perform.

In the 5-fold CV setting (Figure~\ref{fig:5fold}), all models perform significantly better than the \naive model.
Indeed, Figures~\ref{fig:5fold} \cc and \dd show almost perfect agreement between true and predicted probabilities.
Furthermore, we find that ML models can be highly robust to data subsampling (Figure~\ref{fig:5fold}~\ee).
% 100/22*3 = 13.63% ≈ 15%, and 100/22*6 = 27.27% ≈ 25%
Indeed, performance does not meaningfully decrease until less than 25\% of the data is used for training, and good performance is achieved by training on only 15\% of the data.

In leave\-/one\-/cluster\-/out testing (Figure~\ref{fig:leaveclusterout}), the models performed well for most unsampled \ch{Si29H_x} clusters but not for \ch{Si2H6}, likely due to the presence only in the large cluster of low vibrational frequencies that can better accommodate the collision energy.
Since our training set has four \ch{Si29H_x} clusters and only one \ch{Si2H_x} and one \ch{Si4H_x} clusters, it appears that the model is biased towards the behavior of the \ch{Si29H_x} clusters.

Notably, most models showed an increased error for \ch{Si29H18}.
The \ch{Si29H_x} clusters start with the fully saturated \ch{Si29H36} molecule and become less saturated until we get to \ch{Si29H18}.
Therefore, we expect the error to be higher for \ch{Si29H36} because the model has only unsaturated \ch{Si29H_x} clusters to train from, while for the remaining \ch{Si29H_x} clusters, we expected similar errors.
The slightly abnormal behavior of the predictions for collisions involving \ch{Si29H18} suggests interactions dominated by different reaction pathways, possibly related to H isomerization.
The results for the impactors are similar (see \sisec{leaveimpactorout}). 
Overall, the models are most effective when making predictions for similar-sized molecules, at least when provided with such a limited selection.

While the natural conclusion about the need for a wider variety of cluster sizes is correct, it should also be confronted with the fact that not all atomic arrangements are equally stable.
The lower energy associated with specific structures (\eg spherical, truncated polyhedrons) may result in some clustering of the dominant reactive pathways, which may complicate even a model trained on a more varied dataset.

A similar analysis for temperature is shown in Figure~\ref{fig:leavetempout}, where we observe that the model performance is relatively consistent except at the lowest temperature.
The \naive loss is highest for \K{300}, and most models (except DeepSets) perform poorly for \K{300} and even worse for \K{900}.
While we could not determine the reason for this difference beyond the difficulty of extrapolation compared to interpolation (which should affect the \K{900}), it should be noted that physisorption plays a much more significant role at this temperature, hinting again at the model sensitivity to underlying physical and chemical processes. 

As shown in Figure~\ref{fig:stickingsim} and SI Figure~\ref{sifig:stickingtype}, \pst is nearly constant between \K{700} and \K{900}.
Because LGBM learns piecewise-constant functions, it also performs constant extrapolations, which is ideal in this setting.
However, \pst is hardly constant between \K{300} and \K{400}, meaning that constant extrapolations perform poorly.
This is why other methods, such as DeepSets, outperform LGBM when extrapolating to lower temperatures and why LGBM outperforms all other methods but DeepSets when extrapolating to higher temperatures.

\begin{figure}[!htb]
  \centering
  \includegraphics{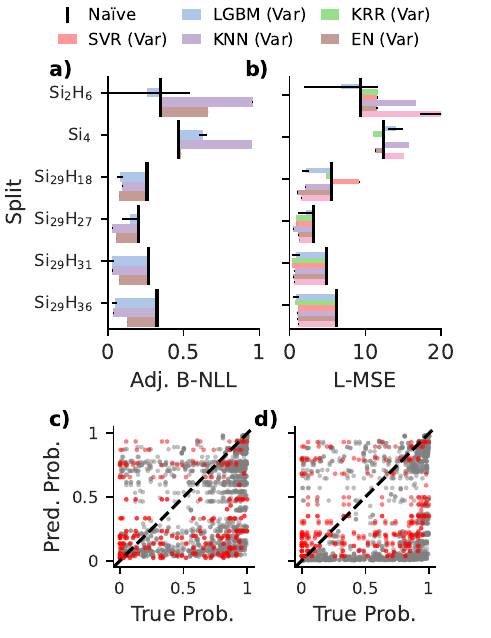}
  \caption{
    Performance of leave\-/one\-/cluster\-/out CV with permutation-variant models.
    \aa and \bb show the average performance of each model trained and evaluated using the adjusted B-NLL and L-MSE, respectively.
    Black bars indicate the standard deviation across random seeds.
    \cc and \dd show the LGBM (Var) predictions for each fold and random seed using the same loss functions as \aa and \bb, respectively.
    Here, we permute the particles before applying the model.
    Predictions for \ch{Si2H6} are highlighted in red.
  }
  \label{fig:leaveclusterout_var}
\end{figure}

\begin{figure}
    \centering
    \includegraphics{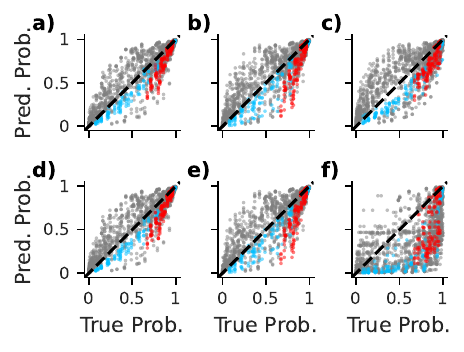}
    \caption{
        \aa--\cc predict \pst for cluster-impactor pairs, while \dd--\ff predict \pst for impactor-cluster pairs.
        \aa, \bb, \dd, and \ee show pseudo-permutation invariant LGBM, while \cc and \ff show standard LGBM.
        \aa, \cc, \dd, and \ff are trained with the Binomial NLL, while \bb and \ee are trained using the LW-MSE.
        Predictions for \ch{SiH3-Si} and \ch{Si2H4} are highlighted in blue and red, respectively.
        All random seeds are shown.
    }
    \label{fig:leaveclusterout_varscatter}
\end{figure}

Notably, we find that permutation-variance significantly impacts the generalization capabilities of all models, as shown in Figure~\ref{fig:leaveclusterout}.
This effect is particularly strong for \ch{Si2H6}, \ch{Si4}, \ch{Si29H27}.
We suspect that this is partial because \ch{Si4} and \ch{Si2H6} are more similar in size to the impactors than the other clusters.
As a result, a permutation-variant model learns that the larger particles are usually on one side, which biases the predictions.
Indeed, comparing Figures~\ref{fig:leaveclusterout} and \ref{fig:leaveclusterout_var}, we find that permutation-variant models achieve surprisingly high and low performance depending on the particle being held out, indicating a tendency to fit and an inability to generalize.
Additionally, panels \cc and \ff in Figure~\ref{fig:leaveclusterout_varscatter} show that permutation-variant models make highly inconsistent predictions for each permutation of clusters and impactors.
Indeed, predictions are inconsistent between permutations and are inaccurate unless the model is trained and tested on the same ordering of clusters and impactors.
This variability demonstrates the importance of permutation-invariance for accurate modeling of \pst in nonthermal plasmas.

Visual inspection of true \vs predicted probabilities shows that the B-NLL provides more robust predictions than the L-MSE.
Both losses achieve good empirical performance, indicating that the L-MSE may be suitable when a binomial NLL loss cannot easily be added to a model.
However, the B-NLL achieves more accurate predictions overall due to its less extreme penalization of outliers.
For example, compare the predictions for LGBM trained with each loss on leave\-/one\-/impactor\-/out cross-validation: Although the worst-performing impactor is equally bad for both B-NLL and L-MSE, the error is lower for intermediately-performing impactors such as \ch{SiH3-SiH} and \ch{Si2H4} when considering the B-NLL (SI Figure~\ref{sifig:leaveimpactorout}).
This is reminiscent of robust regression, where robust losses (\eg Huber loss) are chosen because they do not over-emphasize outliers as significantly as the squared error.
Indeed, the B-NLL loss sits between the L-MSE and logit-transformed Huber Loss (L-H) in Figure~\ref{fig:binom_vs_mse}.
Thus, we recommend using the Binomial NLL when possible, and we suspect that further improvements are possible using robust binomial or logistic regression approaches~\cite{feng2014}.

We find that, in nearly all settings, ML models significantly outperform the \naive prediction.
However, we note that models generally perform best when interpolating (rather than extrapolating) for temperature and that better performance may be achieved at higher temperatures.
Additionally, DeepSets excels at extrapolating to unseen temperatures, and LGBM is the most consistent at successfully extrapolating to unseen structures.
Finally, the loss function and permutation invariance can significantly affect model performance and generalization.
This result suggests that, given a correctly chosen model architecture, we can focus our simulations on a small subset of important conditions (\eg edge temperatures) to derive \pst across many diverse settings effectively.

\section{Conclusions}
\label{sec:conclusion}
Particle growth in a high-energy gas phase, such as during combustion or in nonthermal plasma, is a complex, non-linear process that requires accurate modeling.
Even when narrowing the scope to a specific system and set of reactions, capturing the rates and mechanism remains computationally challenging, even using classical approximations. 
While such detailed descriptions are not always necessary, a more nuanced description of the reaction rates can benefit several contexts, such as more accurate reaction rates, hyper-doping, and core-shell nanoparticle production.
The subtle differences between the various \ch{Si2H_y} species simulated in this paper, as well as \ch{SiH_y}, are symptomatic of a series of competing phenomena (\eg physisorption, energy redistribution) that cannot be easily generalized and that can lead to systematic biases when ignored.

To address this, we have focused on training and testing several permutation-invariant ML models to reduce the computational effort associated with these simulations. Our results show that nearly 90\% of interactions can be predicted using machine learning without significantly impacting accuracy.
Furthermore, we have demonstrated the importance of principled loss functions, model architectures, and sampling procedures for deriving accurate and reliable predictions.
Figure~\ref{fig:5fold} shows that, in general, simple system-specific features are descriptive enough to predict sticking probabilities for silane nanoparticles in nonthermal plasma after training on only a fraction of simulated interactions.
Additionally, Figure~\ref{fig:leavetempout} shows that our model can extrapolate and interpolate quite well for unsampled temperatures.
However, Figure~\ref{fig:leaveclusterout} indicates our specific combination of ML models and input features has difficulty extrapolating to nanoparticles with different degrees of saturation.

These findings demonstrate that ML methods can significantly reduce the computational cost of computing the results of complex reactions in nonthermal plasma and other difficult-to-model systems.
However, careful selection of model architecture and training data is crucial to ensure the generalizability of predictions.
Based on our results, we conclude that the most effective acceleration method involves simulating a subset of particles across a range of temperatures (especially the upper and lower bounds of the expected range), maintaining a balance of relevant molecular properties (\eg H-saturation in this case) in the training set, and then training a permutation-invariant ML model using the binomial negative log-likelihood.

While this study focused on the data collected for a specific system, namely the sticking probability of silanes computed through classical reactive molecular dynamics, the overall approach, both molecular dynamics simulations and analysis of ML models, is relatively general.
As such, we expect that similar methods can be readily adapted to generate more computationally efficient and realistic growth parameters for NTPs, thus improving the efficiency and accuracy of simulations.

\section{Reproducibilty}
\label{sec:reproducibility}
Supporting data and code are currently available through this link:
% \url{https://anonymous.4open.science/r/ntp_silicon-A203}.
\url{https://gitlab.eecs.umich.edu/mattrmd-public/ntp-silicon}.
Data will be provided via a DOI-minting repository upon acceptance.

\section{CRediT authorship contribution statement}

\textbf{Matt Raymond:} Formal analysis, investigation, methodology, software, supervision, validation, visualization, writing - original draft, writing - review \& editing.
\textbf{Paolo Elvati:} Conceptualization, data curation, investigation, methodology, software, supervision, visualization, writing - original draft, writing - review \& editing.
\textbf{Jacob Saldinger:} Conceptualization, data curation, software, supervision, writing - original draft.
\textbf{Jonathan Lin:} Formal analysis, investigation, software, visualization, writing - original draft.
\textbf{Xuetao Shi:} Data curation, software.
\textbf{Angela Violi:} Conceptualization, funding acquisition, project administration, resources, supervision, writing - review \& editing.

\section{Declaration of Competing Interests}
The authors declare that they have no known competing financial interests or personal relationships that could have appeared to influence the work reported in this paper.

\section*{Acknowledgements}
This work has been supported by the US Army Research Office MURI Grant No. W911NF-18-1-0240 and by the NSF ECO-CBET No. F059554.

\twocolumn

\bibliographystyle{iopart-num}
\bibliography{ref}

\appendix
\sinumbering
\clearpage
\onecolumn
\section{Leave Impactor Out}
  \label{sisec:leaveimpactorout}

  \begin{figure*}[h]\centering
    \includegraphics[width=\textwidth]{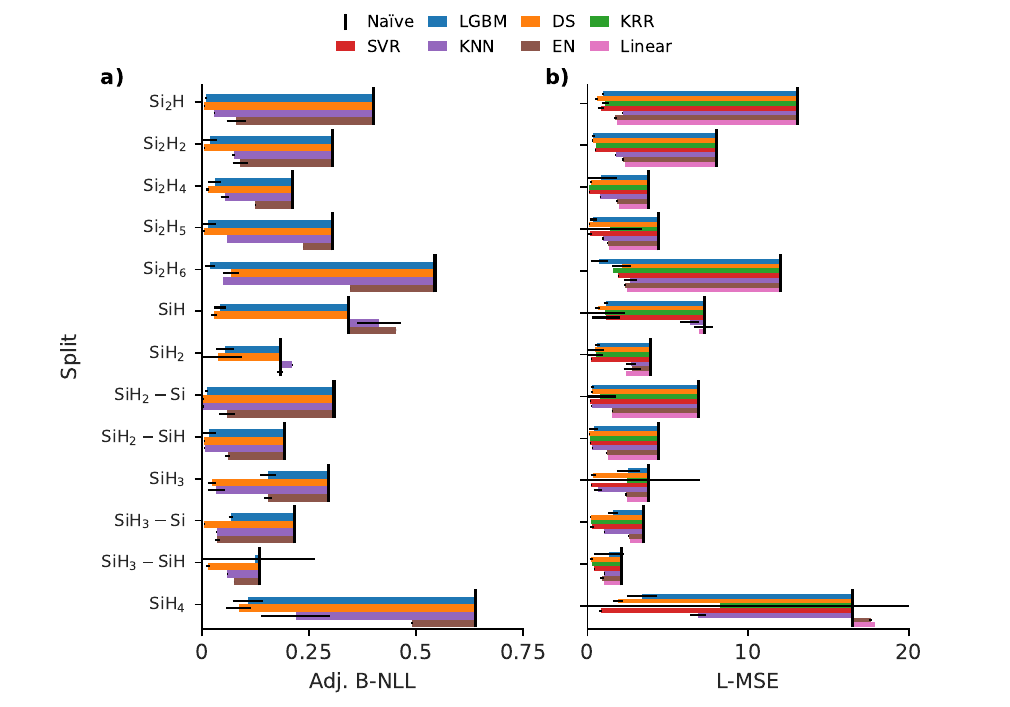}
    \caption{
        Performance of leave\-/one\-/impactor\-/out CV.
        \aa and \bb show the average performance of each model trained and evaluated using the adjusted B-NLL and L-MSE, respectively.
        Black bars indicate the standard deviation across random seeds.
    }
    \label{sifig:leaveimpactorout}
  \end{figure*}

  For leave-one-impactor-out (SI Figure~\ref{sifig:leaveimpactorout}), some models perform especially poorly for \ch{SiH}, \ch{SiH2}, \ch{SiH3-Si}, and \ch{SiH3-SiH}.
  There are several potential explanations for this.
  One is that the \naive prediction error for these particles is already significantly lower than the other particles, leaving less room for improvement.
  For a couple of impactors, namely \ch{SiH4} and \ch{SiH3-SiH3}, most models had nearly double the error as they did for other impactors.
  This may be caused by overfitting due to the large amount of remaining data for partially saturated molecules, suggesting that optimal training data would contain a higher percentage of fully saturated molecules.
  Alternatively, it could simply mean that the sticking probabilities for this molecule are outliers.
  Regardless, DeepSets (and LGBM) significantly outperform the \naive model in all (most) cases.

\section{CV Method}
  \label{sisec:cvmethod}

  In Supplementary Algorithms \ref{sialg:cv_nested} and \ref{sialg:cv_group}, we include the details of the cross-validation methods used in this work.

  \begin{algorithm*}
    \label{sialg:cv_nested}
    \SetAlgoLined
    \SetKwInOut{Input}{Input}
    \SetKwInOut{Output}{Output}

    \Input{Dataset $D$, parameter grid $\mc{G}$, parametric model $f_{\theta, g}$}
    \Output{Performance metric for each outer testing set}

    \tcp{Arrays holding outer test metric}
    $\mu \gets []$\;

    \ForEach{$g \in \mc{G}$}{
      \ForEach{$k \in [5]$}{
        $k' \gets (k+1) \mod 5$\;

        \tcp{Outer cross-validation loop}
        $D^\text{outer}_\text{train} \gets D$ split into 5 folds with the $k$ and $k'$-th folds removed\;
        $D^\text{outer}_\text{test} \gets$ $k$-th fold of $D$\;
        $D^\text{outer}_\text{val} \gets$ $k'$-th fold of $D$\;

        \tcp{Array of losses}
        $\eta_g \gets []$\;

        \ForEach{$j \in [5]$}{
          \tcp{Inner cross-validation loop}
          $D^\text{inner}_\text{train} \gets D^\text{outer}_\text{train}$ split into 5 folds with the $j$ and $j'$-th folds removed\;
          $D^\text{inner}_\text{test} \gets$ $j$-th fold of $D^\text{outer}_\text{train}$\;
          $D^\text{inner}_\text{val} \gets$ $j'$-th fold of $D^\text{outer}_\text{train}$\;

          $\theta^* \gets \arg\min_\theta f_{\theta,g}$ with training and validation sets $D^\text{inner}_\text{train}$ and $D^\text{inner}_\text{train}$\;

          Append the loss computed on $D^\text{inner}_\text{test}$ to $\eta_g$\;
        }

        \tcp{Select the optimal parameters}
        $g^* \gets \arg\min_g \sum\eta_g / 5$\;
        
        \tcp{Use these parameters to retrain the model}
        $\theta^* \gets \arg\min_\theta f_{\theta,g^*}$ with training and validation sets $D^\text{outer}_\text{train}$ and $D^\text{outer}_\text{train}$\;
        Append the loss computed on $D^\text{outer}_\text{test}$ to $\mu$\;
      }
    }

    \tcp{Return array of performance metrics on the test sets}
    \Return{$\mu$}

    \caption{Grid search with nested 5-fold cross-validation for one seed.}
  \end{algorithm*}

  \begin{algorithm*}
    \label{sialg:cv_group}
    \SetAlgoLined
    \SetKwInOut{Input}{Input}
    \SetKwInOut{Output}{Output}

    \Input{Dataset $D$, parameter grid $\mc{G}$, set of groups $\mc{C}$, parametric model $f_{\theta, g}$}
    \Output{Performance metric for each outer testing set}

    \tcp{Arrays holding outer test metric}
    $\mu \gets []$\;

    \ForEach{$g, \in \mc{G}$}{
      \ForEach{$k \in [|\mc{C}|]$}{
        $k' \gets (k+1) \mod |\mc{C}|$\;

        \tcp{Outer cross-validation loop}
        $D^\text{outer}_\text{train} \gets D$ split into groups from $\mc{C}$ with the $k$ and $k'$-th groups removed\;
        $D^\text{outer}_\text{test} \gets$ $k$-th group of $D$\;
        $D^\text{outer}_\text{val} \gets$ $k'$-th group of $D$\;

        \tcp{Array of losses}
        $\eta_g \gets []$\;

        \ForEach{$j \in [|\mc{C}\setminus \set{k,k'}|]$}{
          \tcp{Inner cross-validation loop}
          $D^\text{inner}_\text{train} \gets D^\text{outer}_\text{train}$ split into groups from $\mc{C}\setminus \set{k,k'}$ with the $j$ and $j'$-th groups removed\;
          $D^\text{inner}_\text{test} \gets$ $j$-th group of $D^\text{outer}_\text{train}$\;
          $D^\text{inner}_\text{val} \gets$ $j'$-th group of $D^\text{outer}_\text{train}$\;

          $\theta^* \gets \arg\min_\theta f_{\theta,g}$ with training and validation sets $D^\text{inner}_\text{train}$ and $D^\text{inner}_\text{train}$\;

          Append the loss computed on $D^\text{inner}_\text{test}$ to $\eta_g$\;
        }

        \tcp{Select the optimal parameters}
        $g^* \gets \arg\min_g \sum\eta_g / |\mc{C}\setminus \set{k,k'}|$\;
        
        \tcp{Use these parameters to retrain the model}
        $\theta^* \gets \arg\min_\theta f_{\theta,g^*}$ with training and validation sets $D^\text{outer}_\text{train}$ and $D^\text{outer}_\text{train}$\;
        Append the loss computed on $D^\text{outer}_\text{test}$ to $\mu$\;
      }
    }

    \tcp{Return array of performance metrics on the test sets}
    \Return{$\mu$}

    \caption{
        Grid search with nested leave-x-out cross-validation for one seed.
        In our setting, a ``group'' may be a cluster, impactor, or temperature.
        When $|\mc{C}|$ is large, we instead let $j,j'$ be sets of groups instead of individual groups to reduce runtime.
    }
  \end{algorithm*}

  \section{Grid Search}
  \label{sisec:gridsearch_params}

  Parameters used for model selection in the inner 5-fold cross-validation described in the Methods, Section~\ref{sec:methods_ml}.

  \begin{table}[!ht]
      \centering
      \begin{tabular}{l|l}
            \textbf{Parameter Name} & \textbf{Values} \\
            \hline
            Epochs & 100\,000 \\
            Early stopping epochs & 1\,000 \\
            Activation function & \texttt{relu} \\
            Batch size & 64 \\
            Width & $\set{64,128}$ \\
            Depth of each subnetwork & $\set{2,3}$ \\
            Learning rate & 0.001 \\
            Aggregation function $\rho$ & $\set{\texttt{mean}, \texttt{max}}$\\
            Optimizer & \texttt{adam}
      \end{tabular}
      \caption{Grid search parameters for \textbf{DeepSets}}
  \end{table}
  
  \begin{table}[!ht]
      \centering
      \begin{tabular}{l|l}
            \textbf{Parameter Name} & \textbf{Values} \\
            \hline
            $\alpha$ ($\ell_1,\ell_2$ penalty weight) & $\set{0.0001,0.001,0.01,0.1,1,10}$\\
            $\ell_1$ ratio & $\set{0,0.25, 0.5, 0.75, 1}$\\
            Maximum iterations & 10\,000\,000
      \end{tabular}
      \caption{Grid search parameters for \textbf{ElasticNet} (LW-MSE)}
  \end{table}
    
  \begin{table}[!ht]
      \centering
      \begin{tabular}{l|l}
            \textbf{Parameter Name} & \textbf{Values} \\
            \hline
            $C$ (regularization) & $\set{0.1, 1, 10, 100, 1\,000}$\\
            $\varepsilon$ (tube) & $\set{0.0001, 0.001, 0.01, 0.1, 1}$\\
            $\gamma$ (scale for RBF kernel) & $\set{0.01, 0.1, 1, 10}$
      \end{tabular}
      \caption{Grid search parameters for \textbf{SVR}}
  \end{table}
  
  \begin{table}[!ht]
      \centering
      \begin{tabular}{l|l}
            \textbf{Parameter Name} & \textbf{Values} \\
            \hline
            $\alpha$ (regularization) & $\set{0.01, 0.1, 1, 10, 100}$\\
            $\gamma$ (scale for RBF kernel) & $\set{0.01, 0.1, 1, 10, 100}$
      \end{tabular}
      \caption{Grid search parameters for \textbf{KRR}}
  \end{table}
    
  \begin{table}[!ht]
      \centering
      \begin{tabular}{l|l}
            \textbf{Parameter Name} & \textbf{Values} \\
            \hline
            Neighbors & $\set{3,6,9}$ \\
            Weighting method & $\set{\texttt{uniform}, \texttt{distance}}$\\
            $p$ (for the $\ell_p$-norm distance) & $\set{1,2}$\\
      \end{tabular}
      \caption{Grid search parameters for \textbf{KNN}}
  \end{table}
  
  \begin{table}[!ht]
      \centering
      \begin{tabular}{l|l}
            \textbf{Parameter Name} & \textbf{Values} \\
            \hline
            Number of estimators & $\set{100,1\,000}$ \\
            $alpha$ ($\ell_1$ regularization) & $\set{0,0.1,1}$\\
            $\lambda$ ($\ell_2$ regularization) & $\set{0, 0.1, 1.0}$\\
      \end{tabular}
      \caption{Grid search parameters for \textbf{LGBM}}
  \end{table}

    \begin{table}[!ht]
      \centering
      \begin{tabular}{l|l}
            \textbf{Parameter Name} & \textbf{Values} \\
            \hline
            $C$ (inverse of regularization strength) & $\set{0.01, 0.1, 1.0}$\\
            $\ell_1$ ratio & $\set{0, 0.25, 0.5, 0.75, 1}$\\
            Solver & \texttt{saga}\\
            Penalty & \texttt{elasticnet}\\
      \end{tabular}
      \caption{Grid search parameters for \textbf{ElasticNet} (Binomial loss via Logistic Regression)}
    \end{table}

\newpage
\section{MD additional results}

\begin{figure}[!htb]
  \centering
  \includegraphics[width=0.9\linewidth]{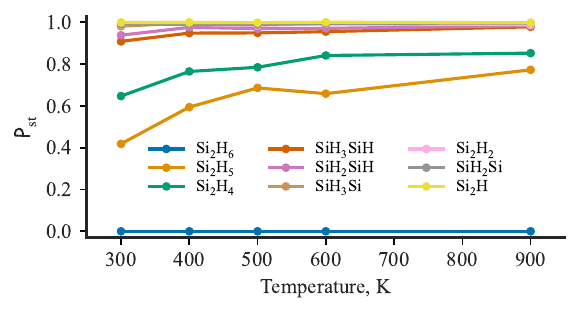}
  \caption{
    Fraction of chemisorption sticking events between \ch{Si29H36} and different \ch{Si2H_x} silicon fragments at various temperatures.
  }
  \label{sifig:stickingtype}
\end{figure}

\newpage
  \section{True \vs Predicted values}

  Here, we plot each cross-validation procedure's true \vs predicted sticking probability.
  We plot the test points for every cross-validation loop that used the selected parameters.

  \begin{figure*}
      \centering
      \includegraphics[width=0.75\textwidth]{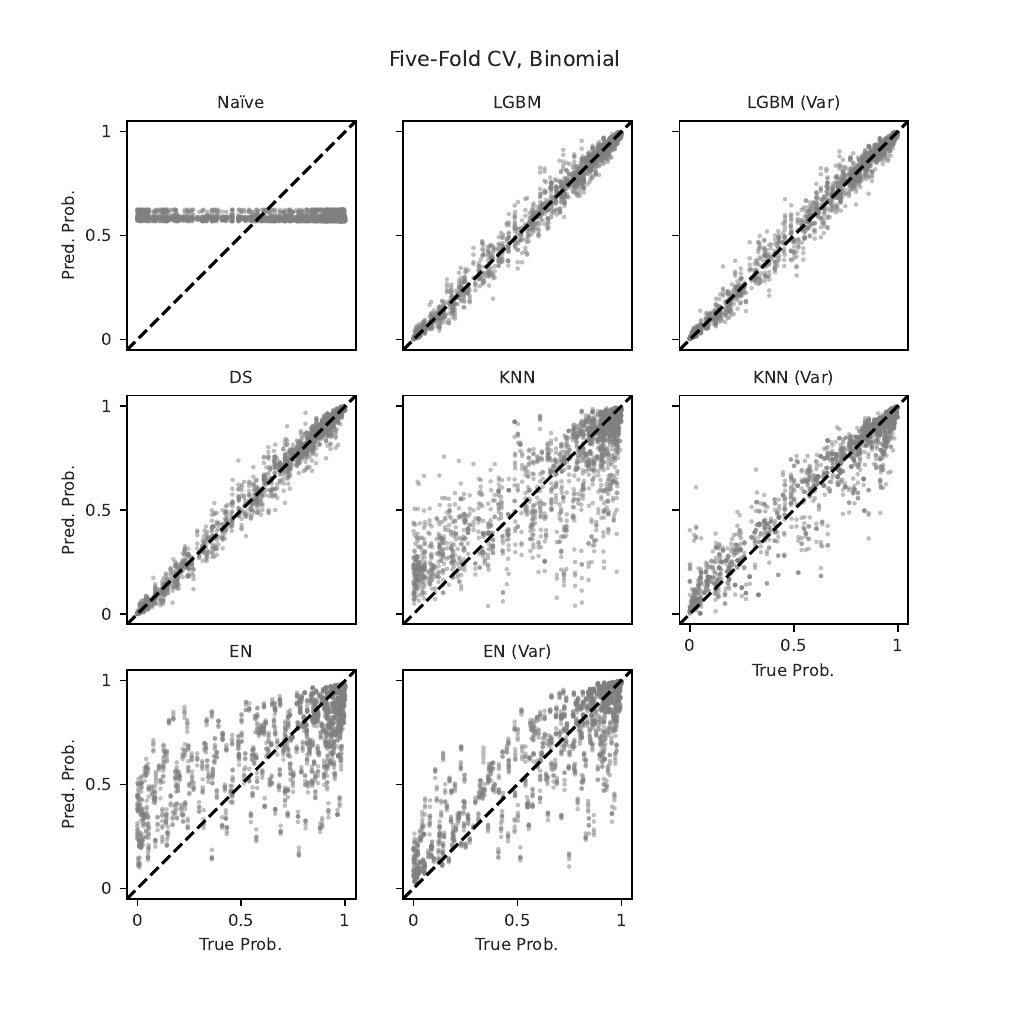}
      \caption{
        True \vs predicted probabilities for each model using nested \textbf{five-fold} cross-validation and the Binomial loss.
      }
      \label{sifig:5fold_scatter_binom}
  \end{figure*}
  
  \begin{figure*}
      \centering
      \includegraphics[width=0.75\textwidth]{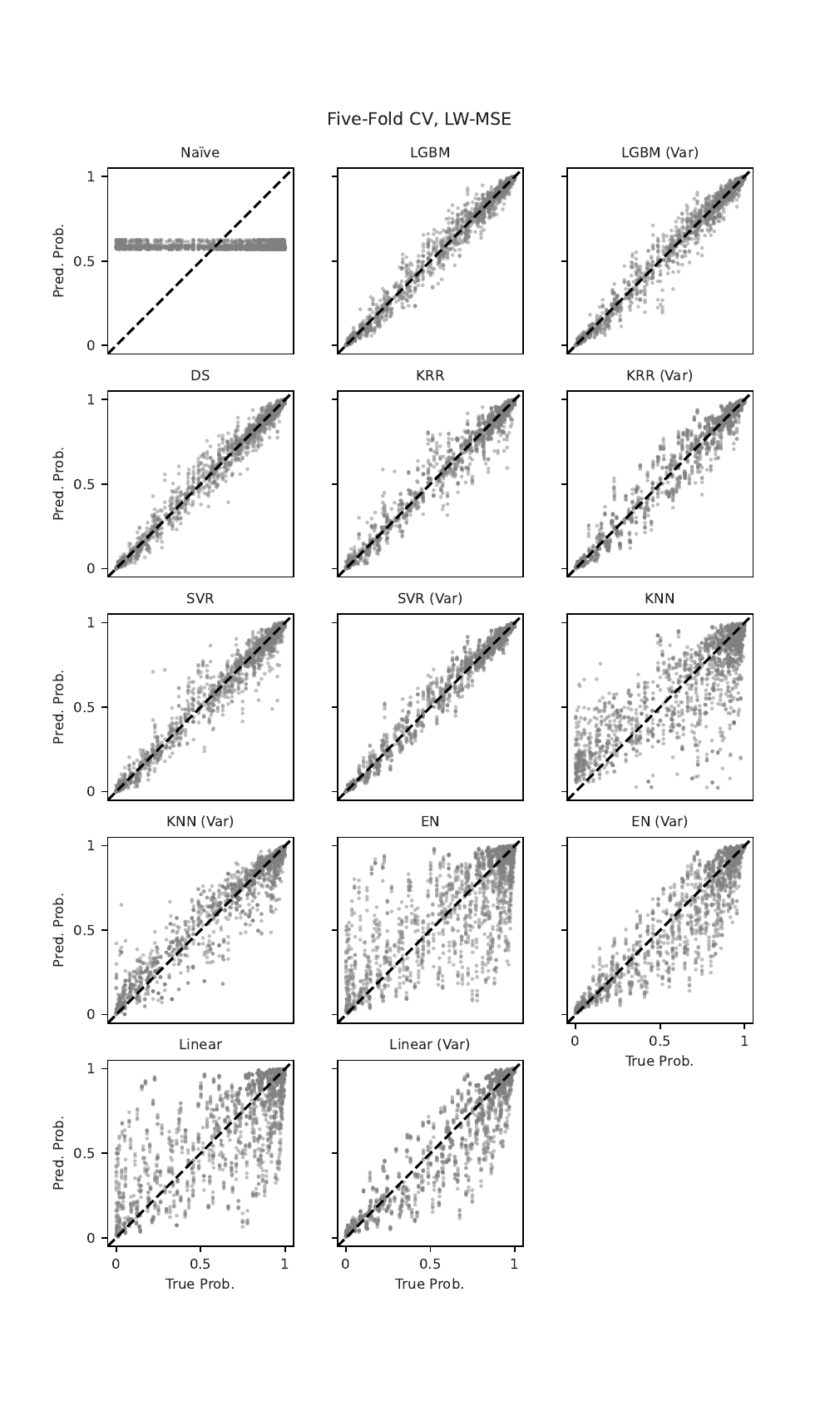}
      \caption{
        True \vs predicted probabilities for each model using nested \textbf{five-fold} cross-validation and the LW-MSE loss.
      }
      \label{sifig:5fold_scatter_mse}
  \end{figure*}

  \begin{figure*}
      \centering
      \includegraphics[width=0.75\textwidth]{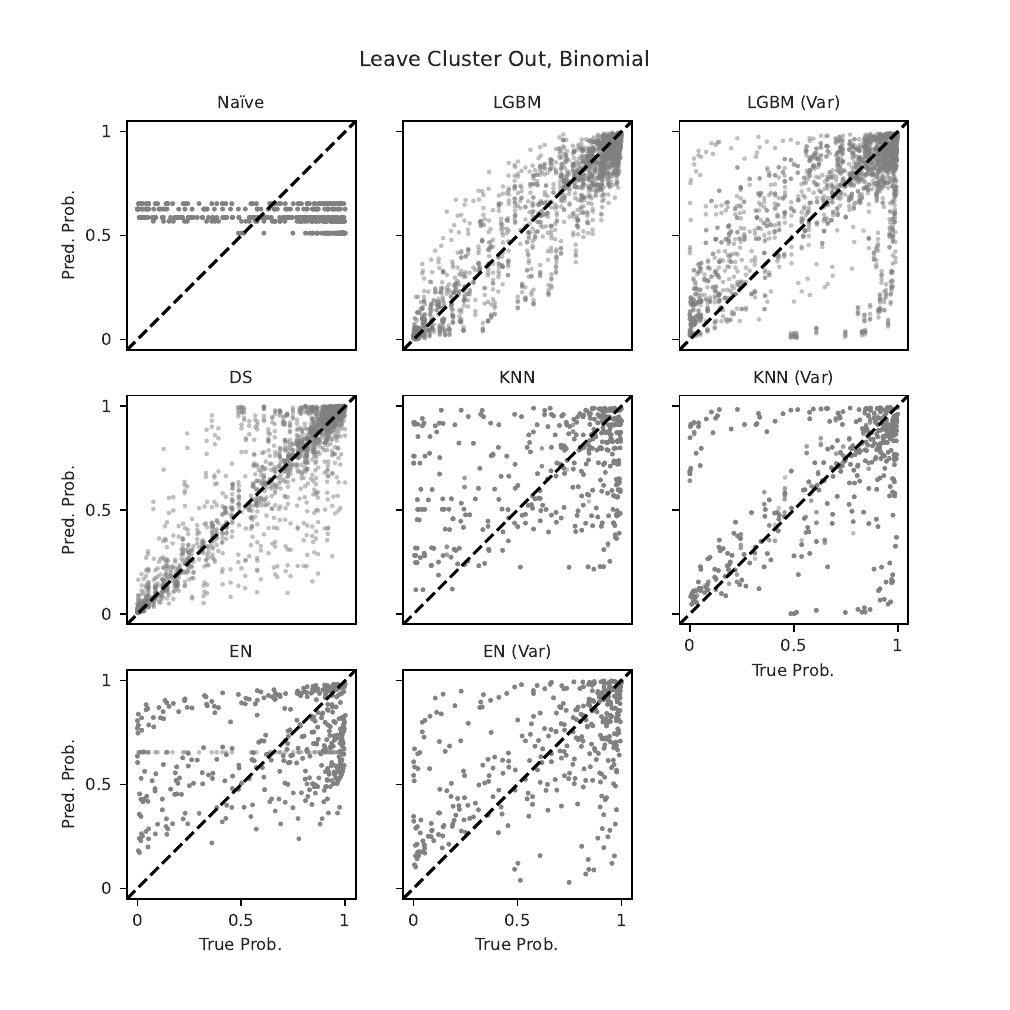}
      \caption{
        True \vs predicted probabilities for each model using nested \textbf{leave-one-cluster-out} cross-validation and the Binomial loss.
      }
      \label{sifig:cluster_out_scatter_binom}
  \end{figure*}
  
  \begin{figure*}
      \centering
      \includegraphics[width=0.75\textwidth]{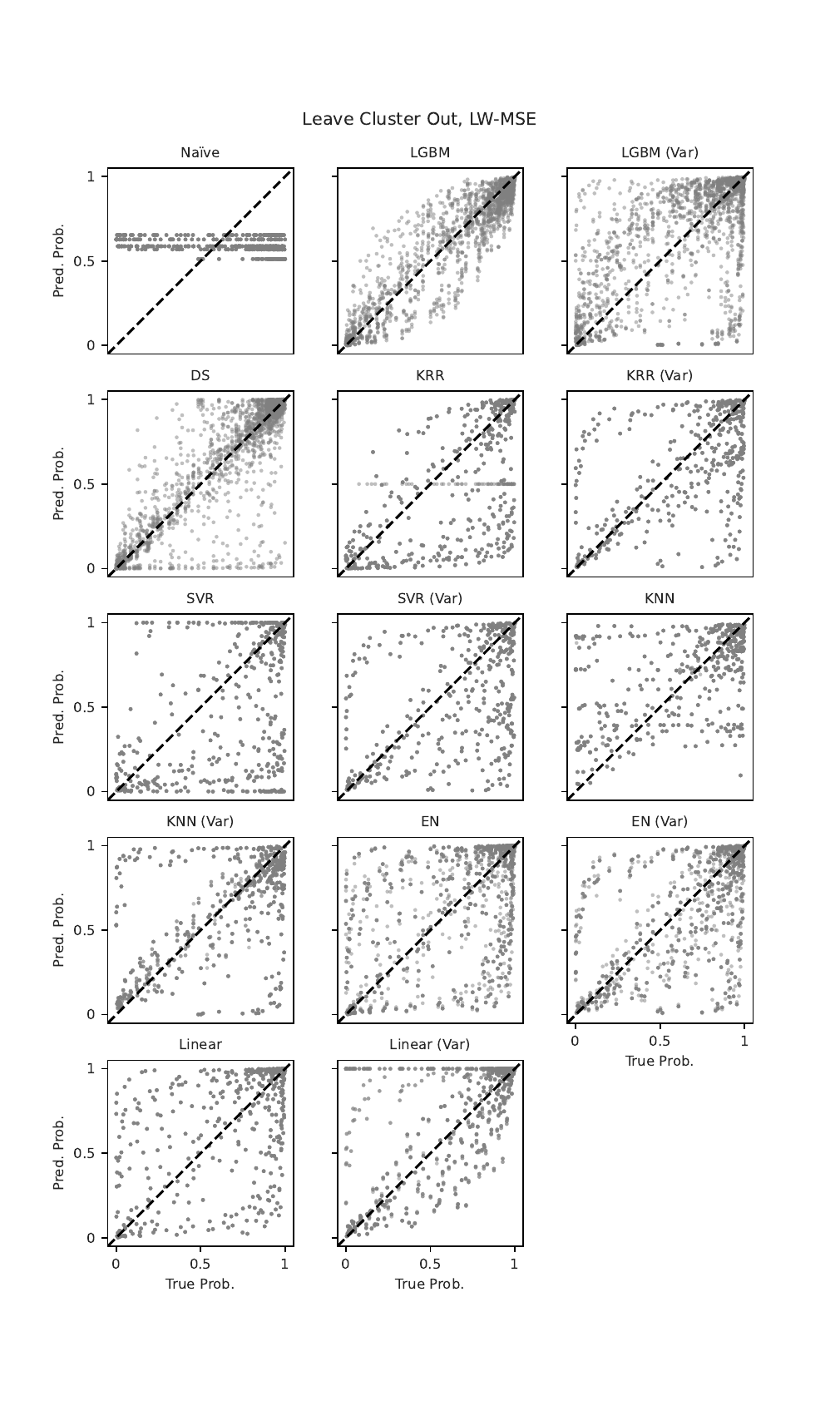}
      \caption{
        True \vs predicted probabilities for each model using nested \textbf{leave-one-cluster-out} cross-validation and the LW-MSE loss.
      }
      \label{sifig:cluster_out_scatter_mse}
  \end{figure*}

  \begin{figure*}
      \centering
      \includegraphics[width=0.75\textwidth]{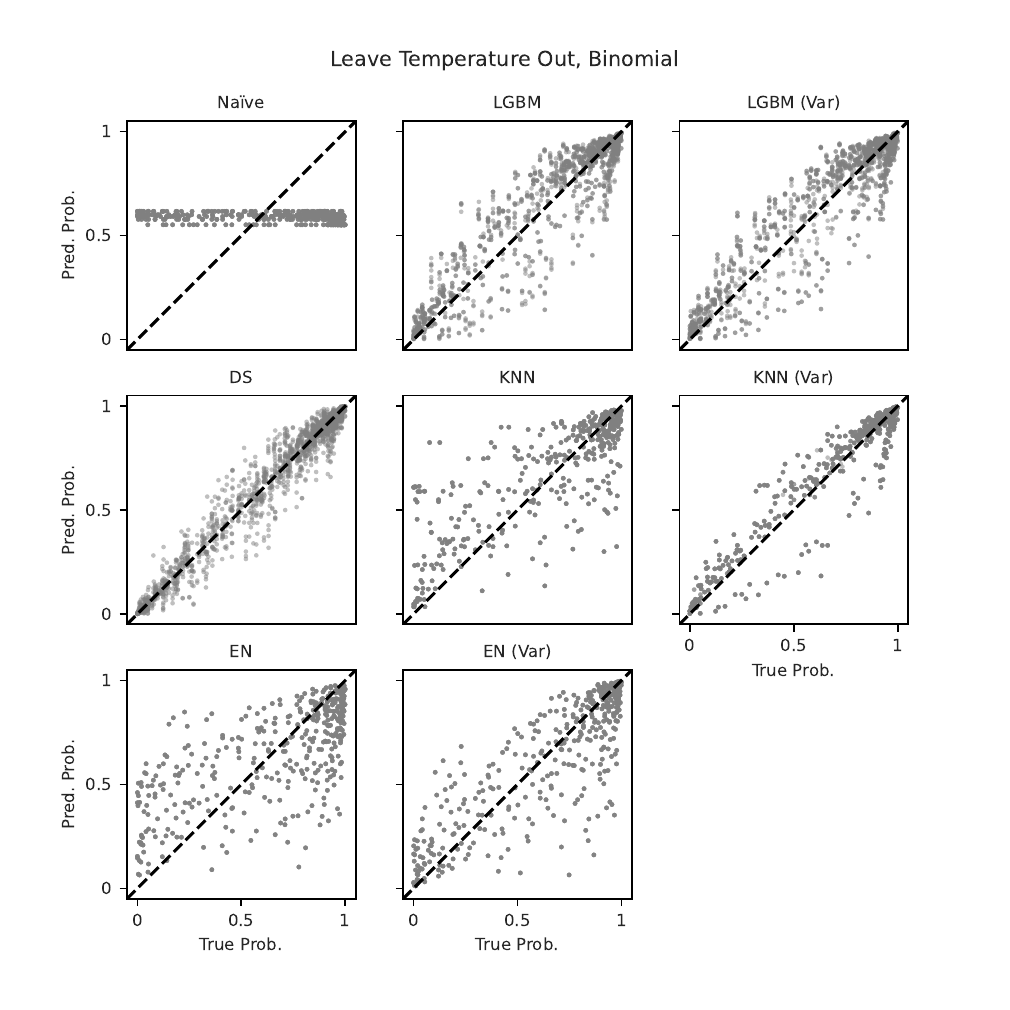}
      \caption{
        True \vs predicted probabilities for each model using nested \textbf{leave-one-temperature-out} cross-validation and the Binomial loss.
      }
      \label{sifig:temp_out_scatter_binom}
  \end{figure*}
  
  \begin{figure*}
      \centering
      \includegraphics[width=0.75\textwidth]{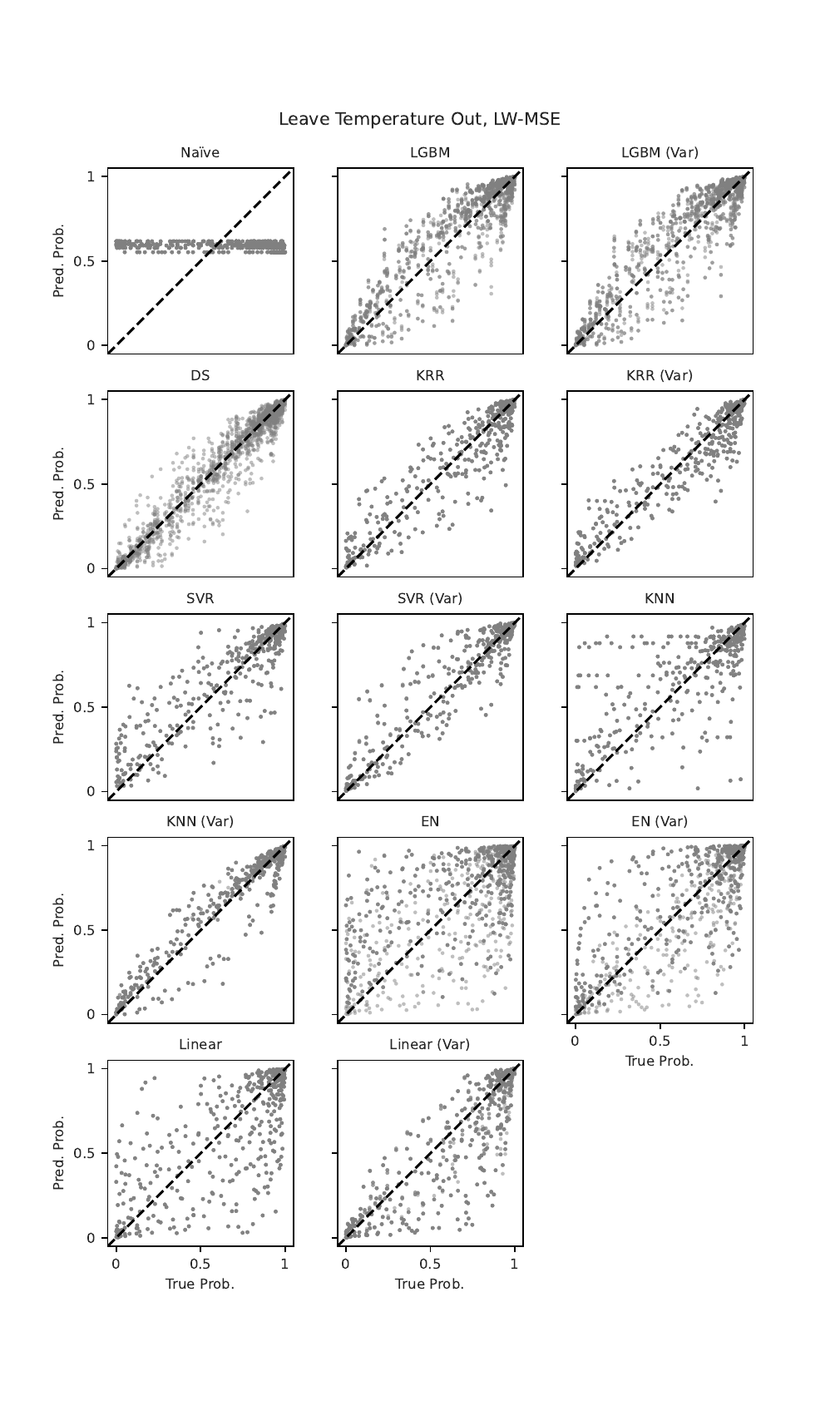}
      \caption{
        True \vs predicted probabilities for each model using nested \textbf{leave-one-temperature-out} cross-validation and the LW-MSE loss.
      }
      \label{sifig:temp_out_scatter_mse}
  \end{figure*}
  
  \begin{figure*}
      \centering
      \includegraphics[width=0.75\textwidth]{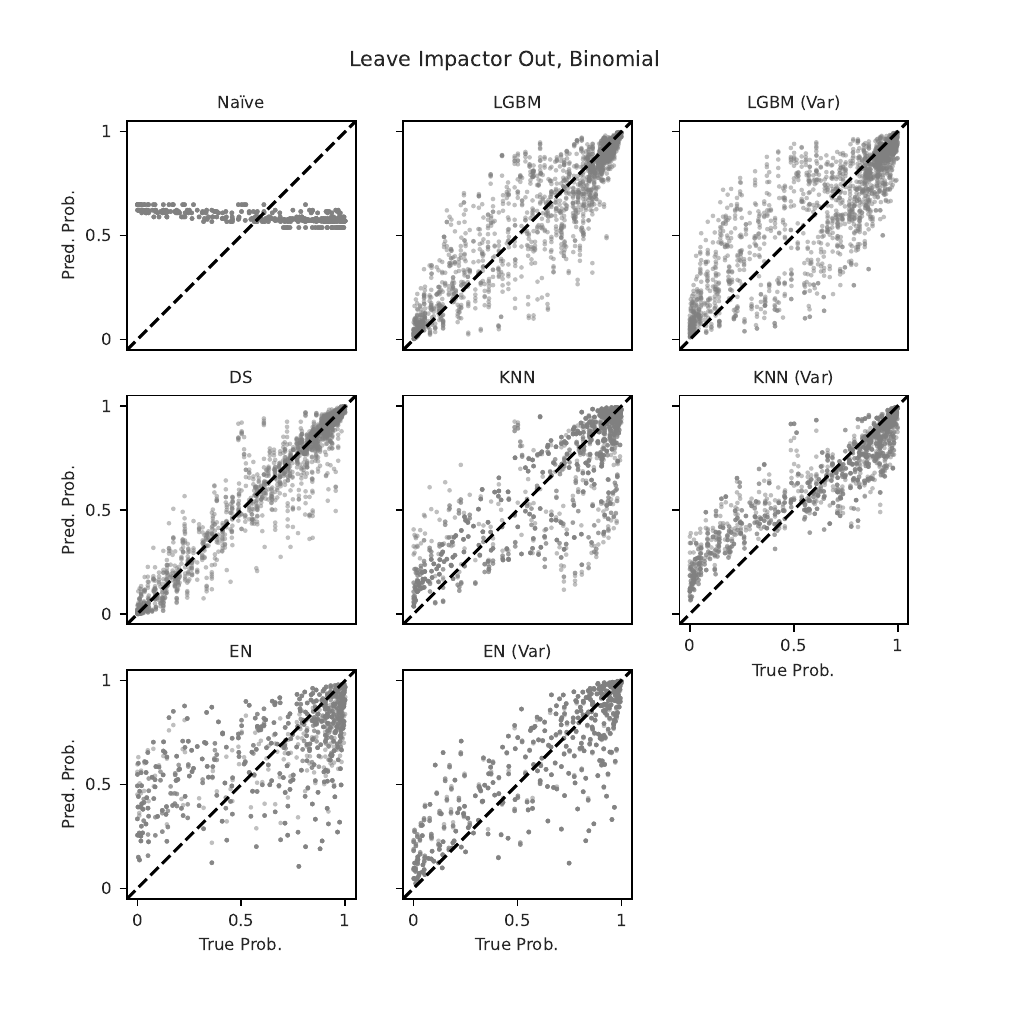}
      \caption{
        True \vs predicted probabilities for each model using nested \textbf{leave-one-impactor-out} cross-validation and the Binomial loss.
      }
      \label{sifig:impactor_out_scatter_binom}
  \end{figure*}
  
  \begin{figure*}
      \centering
      \includegraphics[width=0.75\textwidth]{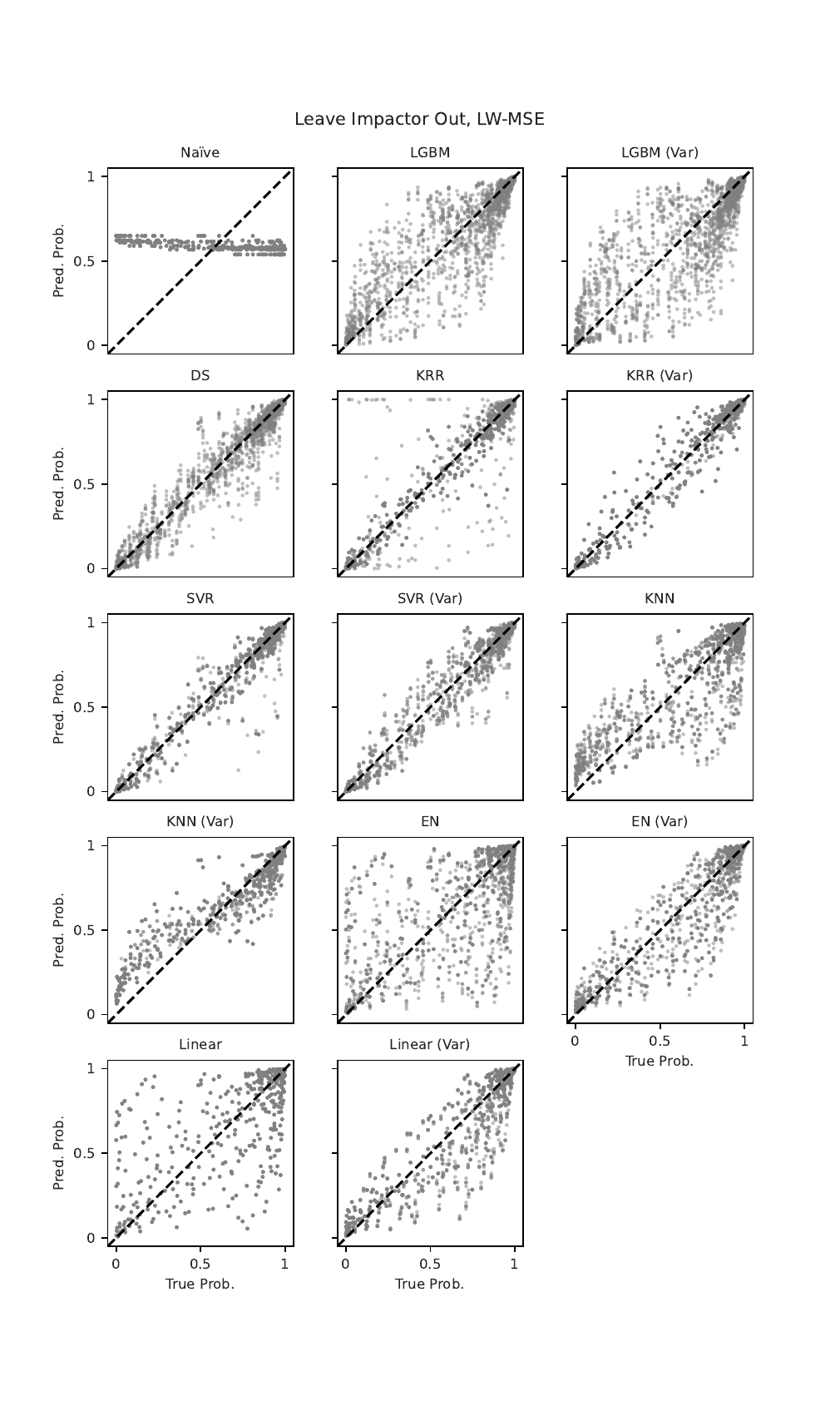}
      \caption{
        True \vs predicted probabilities for each model using nested \textbf{leave-one-impactor-out} cross-validation and the LW-MSE loss.
      }
      \label{sifig:impactor_out_scatter_mse}
  \end{figure*}

  \section{Example Input and Output}

  \begin{table*}[!ht]
      \centering
          \begin{tabular}{ccccccccc|c}
             \multicolumn{4}{c}{\textbf{Cluster Descriptors}} & \multicolumn{4}{c}{\textbf{Impactor Descriptors}} & \textbf{Environmental Descriptors} & \textbf{Outputs}\\
             \#Si & \#h & \#e$_1$ & \#e$_2$ & \#Si & \#h & \#e$_1$ & \#e$_2$ & Temperature (K) & Probability \\
             \hline
             4 & 0 & 8 & 0 & 1 & 1 & 3 & 0 & 300 & 0.979 \\
             2 & 6 & 0 & 0 & 1 & 4 & 0 & 0 & 600 & 0.000 \\
             \vdots & \vdots & \vdots & \vdots & \vdots & \vdots & \vdots & \vdots & \vdots & \vdots
        \end{tabular}
      \caption{
        Example inputs and outputs for the machine learning models (assuming no pre-processing).
        \# Si and \# h indicate the number of silicon and hydrogen atoms, respectively.
        \# e$_1$ and \# e$_2$ indicate the first and second elements in the vector $\#\textbf{e}$ describing the number of unpaired electrons per silicon atom.
      }
      \label{tab:my_label}
  \end{table*}
\end{document}